\documentclass[aps,prl,twocolumn,raggedbottom, reprint]{revtex4-2}

\usepackage[usenames,dvipsnames]{xcolor}
\usepackage{amssymb} 
\usepackage{amsmath}
\usepackage{amsthm}
\usepackage{mathtools}
\usepackage{amsfonts} 
\usepackage{mathrsfs}
\usepackage{hyperref}
\usepackage{graphicx}
\usepackage{mciteplus}
\usepackage{skak}
\usepackage{empheq}
\usepackage{tikz}
\usepackage{booktabs}
\usepackage{siunitx}
\usepackage{makecell}
\usepackage{float}
\usepackage{dsfont}
\usepackage{bbm}

\usepackage[most]{tcolorbox}

\usepackage{braket}
\usepackage{subcaption}

\DeclareFontFamily{OT1}{pzc}{}
\DeclareFontShape{OT1}{pzc}{m}{it}{<-> s * [1.10] pzcmi7t}{}
\DeclareMathAlphabet{\mathpzc}{OT1}{pzc}{m}{it}

\def\be#1\ee{\begin{align}#1\end{align}}

\makeatletter
\newcommand{\bdryno}{\mathpalette\bdry@no\relax}
\newcommand{\bdry@no}[2]{%
  \mspace{1mu}%
  \vbox{%
    \hbox{$\m@th#1\scriptstyle{\ast}$}
    \nointerlineskip
    \kern.25ex
    \hbox{$\m@th#1\scriptstyle{\ast}$}
    \kern-.06ex
  }%
  \mspace{1mu}%
}
\makeatother

\hypersetup{
    pdfencoding=unicode,
	colorlinks=true,
	urlcolor=Maroon,
	linkcolor=RoyalBlue,
	citecolor=Maroon,
	pdftitle={The Complex Liouville String},
	pdfauthor={Scott Collier, Lorenz Eberhardt, Beatrix M\"uhlmann, Victor Rodriguez},
	pdfdisplaydoctitle=true,
	pdfstartview=FitH,
	linktocpage=true
}

\usetikzlibrary{arrows}
\usetikzlibrary{decorations.pathmorphing}
\usetikzlibrary{ decorations.markings}
\usetikzlibrary{shapes.misc}
\tikzset{cross/.style={cross out, draw=black, minimum size=2*(#1-\pgflinewidth), inner sep=0pt, outer sep=0pt},
cross/.default={1pt}}
\usetikzlibrary{shapes.geometric}
\tikzset{
    partial ellipse/.style args={#1:#2:#3}{
        insert path={+ (#1:#3) arc (#1:#2:#3)}
    }
}

\definecolor{bleudefrance}{rgb}{0.19, 0.55, 0.91}
\definecolor{candyapplered}{rgb}{1.0, 0.03, 0.0}
\definecolor{myblue}{rgb}{0.0,0.635,1}
\definecolor{vert}{rgb}{0.1367 0.543 0.1367}

\newcommand{\xx}{\mathsf{x}}
\newcommand{\yy}{\mathsf{y}}

\newcommand{\bM}{\overline{\mathcal{M}}}

\renewcommand\d{\text{d}}

\newcommand\CC{\mathbb{C}}
\newcommand\ZZ{\mathbb{Z}}
\newcommand\RR{\mathbb{R}}

\DeclareMathOperator\tr{tr}

\begin{document}

\begin{flushright}
    \hfill{\tt MIT-CTP/5776}
\end{flushright}

\title{The complex Liouville string}

\author{Scott Collier}

\affiliation{Center for Theoretical Physics, Massachusetts Institute of Technology, Cambridge, MA 02139, USA}

\author{Lorenz Eberhardt}
\affiliation{Institute for Theoretical Physics, University of Amsterdam, 1098 XH Amsterdam, Netherlands}

\author{Beatrix M\"uhlmann}
\affiliation{School of Natural Sciences, Institute for Advanced Study, Princeton, NJ 08540, USA}
\affiliation{Department of Physics, McGill University, Montr\'eal, QC, H3A 2T8, Canada}

\author{Victor A. Rodriguez}
\affiliation{Department of Physics, University of California, Santa Barbara, CA 93106, USA}
\affiliation{Joseph Henry Laboratories, Princeton University, Princeton, NJ 08544, USA}

\date{\today}

\begin{abstract}\noindent
    We introduce the complex Liouville string, a solvable string theory defined by coupling two Liouville theories with complex conjugate central charges $c \in 13+i \RR$ on the worldsheet. We compute its amplitudes from first principles and establish a duality with a double-scaled two-matrix integral. We also analyze general worldsheet boundaries and non-perturbative effects in the genus expansion. By expressing the complex Liouville string as a 2d dilaton gravity theory with a sine potential, we show that it admits both AdS$_2$ and dS$_2$ vacua. 
\end{abstract}

\maketitle

\section{Introduction}  

Solvable yet interesting models of quantum gravity are rare, but yield valuable insights.
A common paradigm for two-dimensional theories of gravity is to define them as a critical worldsheet string theory, allowing for the application of extensive technical machinery. Notable examples include the $(p,q)$-minimal string \cite{Gross:1989vs, Douglas:1989ve, Brezin:1990rb, Seiberg:2004at} or the Virasoro minimal string \cite{Collier:2023cyw}, both of which admit a dual description in terms of double-scaled matrix integrals---providing a version of holographic duality for these theories.
The presentation as a worldsheet string theory, or as a matrix integral, also provides a non-perturbative definition of the corresponding 2d model of quantum gravity. 

These string theories can also be viewed as more traditional theories of gravity, consisting of a metric coupled to a dilaton scalar field. In this formulation, the holographic duality was reinterpreted as a duality between `pure' quantum gravity and a random matrix model \cite{Saad:2019lba}. 

In this letter, we announce a new such model which is far richer than previous ones, but remains solvable.
On the worldsheet, it is defined by coupling two Liouville theories with complex central charges,
\begin{equation}
\begin{array}{c}
\text{Liouville CFT} \\ \text{$c^+\in  13+i\mathbb{R}_+$}
\end{array}
\, \oplus\,  
\begin{array}{c} (\text{Liouville CFT})^* \\ \text{$c^- \in  13 - i\mathbb{R}_+$} \end{array} \, \oplus\,  
\begin{array}{c} \text{$\mathfrak{b}\mathfrak{c}$-ghosts} \\ \text{$c= -26$} \end{array}\, .
\label{eq:Liouville squared worldsheet definition}
\end{equation}
We refer to this theory as the complex Liouville string ($\CC$LS). This theory is very interesting from a variety of perspectives and holds a surprising number of lessons in stock. The string amplitudes can be computed exactly from first principles by employing analytic bootstrap techniques.
It admits a dual description in terms of a random two-matrix model, and moreover describes a theory of 2d quantum gravity with vacua of both positive and negative cosmological constants. Thus it provides an exact version of holography involving de Sitter (dS) spacetimes, which is reflected in several unusual properties of the worldsheet theory and of the matrix model. The theory allows us to consider observables with general boundaries from the worldsheet, and to study (doubly) non-perturbative effects in the sum over topologies in detail.
In this letter, we give an overview of these aspects of $\CC$LS and refer to the papers \cite{paper1, paper2, paper3, paper4} for further details.

\section{Worldsheet theory} \label{sec:worldsheet}
\paragraph{Definition.} 
The worldsheet theory is defined by coupling two complex conjugate Liouville CFTs. We employ the standard parameterization $c= 1+6(b+b^{-1})^2$ of the central charge \cite{Seiberg:1990eb, Dorn:1994xn, Zamolodchikov:1995aa, Teschner:1995yf}. $c^\pm \in 13+i \RR$ in (\ref{eq:Liouville squared worldsheet definition}) corresponds to $b^- = - i b^+ \in \mathrm{e}^{-\frac{\pi i}{4}}\mathbb{R}$.
The combined worldsheet theory can be endowed with a non-unitary inner product via the reality condition $(L_n^+)^\dag=L_{-n}^-$ where $L_n^\pm$ are the Virasoro generators of the two Liouville theories.
Vertex operators in the full worldsheet theory can be written as a product of the Liouville vertex operators $V_{p^+}^+V_{p^-}^-$, with conformal weights given by $h^{\pm} = (c^{\pm}-1)/24-(p^{\pm})^2$ for each primary. The reality condition implies $h^-= (h^+)^*$, while the mass-shell condition imposes $h^++h^-=1$. This is equivalent to $p^- = ip^+$, $p^+\in \mathrm{e}^{-\frac{\pi i}{4}}\mathbb{R}_+$. On-shell vertex operators of $\mathbb{C}$LS including the ghosts are thus $\mathcal{V}_p \equiv \mathcal{N}_b(p)\mathfrak{c}\tilde{\mathfrak{c}}V_{p}^+V_{ip}^-$, where $\mathcal{N}_b(p)$ reflects an arbitrary normalization.
We can then define perturbative string amplitudes as 
 \be \label{eq:definition Agn}
\mathsf{A}_{g,n}^{(b)}(\boldsymbol{p})\equiv C_{\Sigma_g}^{(b)} \int_{\mathcal{M}_{g,n}} \bigg\langle \prod_{k=1}^{3g-3+n} \mathcal{B}_k \widetilde{\mathcal{B}}_k \prod_{j=1}^n \mathcal{V}_{p_j} \bigg \rangle_{\Sigma_{g,n}} ~,
\ee
where bold letters refer to tuples $\boldsymbol{p}=(p_1,\dots,p_n)$. 
$\mathcal{M}_{g,n}$ denotes the moduli space of a genus $g$ Riemann surface with $n$ punctures and $\mathcal{B}_k$ denotes the $3g-3+n$ zero modes of the $\mathfrak{b}$-ghost.
The correlation functions in (\ref{eq:definition Agn}) can be obtained from a conformal block expansion of Liouville theory, which is solved via the DOZZ formula for the structure constants \cite{Dorn:1994xn,Zamolodchikov:1995aa}.
Contrary to the standard bosonic string in 26 dimensions, the moduli space integral in \eqref{eq:definition Agn} is absolutely convergent in $\mathbb{C}$LS.

The full string amplitudes are obtained by summing over genera with a string coupling constant $g_\text{s}=\mathrm{e}^{-S_0}$
\be 
\mathsf{A}_n^{(b)}(S_0;\boldsymbol{p})=\sum_{g \ge 0} \mathrm{e}^{S_0(2-2g-n)} \mathsf{A}_{g,n}^{(b)}(\boldsymbol{p})~. \label{eq:sum over genera}
\ee

\paragraph{Analyticity.} The amplitudes \eqref{eq:definition Agn} are intially defined for $p_j \in \mathrm{e}^{-\frac{\pi i}{4}} \RR_+$ and $b \in \mathrm{e}^{\frac{\pi i}{4}} \RR_+$, but may be analytically continued. 
On a certain open neighborhood of this real locus, the integral over $\mathcal{M}_{g,n}$ is still absolutely convergent and thus $\mathsf{A}_{g,n}^{(b)}(\boldsymbol{p})$ is an analytic function there. Eventually, however, the Liouville correlators develop resonance poles when analytically continued outside the physical region \cite{Zamolodchikov:1995aa, Teschner:2001rv}; this happens for
\be 
\sum_{j=1}^n \pm p_j =rb+s b^{-1}~,\ |r|,\, |s| \in \ZZ_{\ge 0}+g-1+\frac{n}{2} \label{eq:general pole locations}
\ee
for any choice of signs. These poles carry over to the string amplitudes $\mathsf{A}_{g,n}^{(b)}$.
Furthermore, the Liouville correlators can pick up discrete terms from poles crossing the OPE contour of the conformal block expansion under analytic continuation. These spoil convergence over $\mathcal{M}_{g,n}$; and while $\mathsf{A}_{g,n}^{(b)}(\boldsymbol{p})$ can be analytically continued further, the string amplitude $\mathsf{A}_{g,n}^{(b)}(\boldsymbol{p})$ develops branch cuts. 
The discontinuity through the cut is computed from a local analysis near the corresponding degeneration of the worldsheet and can be expressed in terms of string amplitudes with lower values of $3g-3+n$.

\paragraph{Symmetries.} The definition \eqref{eq:definition Agn} has a number of obvious symmetries stemming from the ambiguity of the parameterization of the central charge through $b$, interchanging of the theories, and reflection symmetry of the vertex operators, which in our normalization read 
\begin{subequations}
\begin{align}
    \mathsf{A}_{g,n}^{(b^{-1})}(\boldsymbol{p})&= (-1)^n \mathsf{A}_{g,n}^{(b)}(\boldsymbol{p})\quad   \text{(duality)}~, \label{eq:Agn duality symmetry} \\
    \mathsf{A}_{g,n}^{(-ib)}(i\boldsymbol{p}) &= (-i)^n \mathsf{A}_{g,n}^{(b)}(\boldsymbol{p})\quad\quad  \text{(swap)}~, \label{eq:swap symmetry}\\
    \mathsf{A}_{g,n}^{(b)}(-p_1,p_2,\dots,p_n) &= - \mathsf{A}_{g,n}^{(b)}(\boldsymbol{p})\quad ~~\, \text{(reflection)}~, \label{eq:reflection symmetry}
\end{align} \label{eq:Agn symmetries}%
\end{subequations}
as well as permutation symmetry in the Liouville momenta. The perhaps unusual minus sign in \eqref{eq:reflection symmetry} follows from a convenient normalization. 

\paragraph{Dilaton equation and higher equations of motion.} 
Further constraints on the string amplitudes can be derived when specializing one of the momenta to a degenerate value $p_j=\frac{1}{2}(m b-n b^{-1})$, $m,n \in \ZZ$. One can use the technology of higher equations of motion developed in the context of minimal string theory \cite{Zamolodchikov:2003yb,Belavin:2006ex} to write the integrand in \eqref{eq:definition Agn} as a total derivative and localize it to a sublocus in $\mathcal{M}_{g,n}$. 
The simplest such relation is the dilaton equation,
\be 
\lim_{p' \to \frac{1}{2}\widehat{Q}} \mathsf{A}_{g,n+1}^{(b)}(\boldsymbol{p},p')=\Big(\tfrac{Q\chi_{g,n}}{2}+\sum_{j=1}^n \sqrt{p_j^2}\, \Big)\mathsf{A}_{g,n}^{(b)}(\boldsymbol{p})~,\label{eq:dilaton equation}
\ee
with $Q=b^{-1}+b$, $\widehat{Q}=b^{-1}-b$ and $\chi_{g,n}=2-2g-n$. The square root appears on the right hand side since the string amplitude is analytically continued outside its initial domain of definition where it develops branch cuts as discussed above. 

\paragraph{Analytic bootstrap.} Except for $\mathsf{A}_{0,3}^{(b)}(p_1,p_2,p_3)$ which does not involve moduli integration, it is challenging to evaluate \eqref{eq:definition Agn} from first principles since both $\mathcal{M}_{g,n}$ as well as the Virasoro conformal blocks that are necessary to define the worldsheet correlators are very complicated. 
Nonetheless, $\mathsf{A}_{g,n}^{(b)}$ can be inferred from the various properties mentioned above. This constitutes an analytic bootstrap problem. For $(g,n)=(0,4)$ and $(g,n)=(1,1)$, it has a unique solution, provided one assumes subexponential growth on $\mathsf{A}_{g,n}^{(b)}(\boldsymbol{p})$ for large values of $|p_j|$, which we were not able to establish analytically. We will present the result below, see eq.~\eqref{eq:Agn stable graphs}.
For larger values of $(g,n)$, we have fewer constraints and the bootstrap problem becomes more complex. Therefore we did not try to solve it exhaustively, but we will check below that the general conjecture \eqref{eq:Agn stable graphs} is consistent with all the constraints that we derived from the worldsheet. For $(g,n) = (0,4)$ and $(1,1)$ we can evaluate the string amplitudes by direct numerical integration over moduli space, finding results  consistent with the analytic formula within $<10^{-4} \%$.

\section{Matrix model} \label{sec:matrix model}
The outputs of the analytic bootstrap are remarkably compact formulas for the string amplitudes whose underlying simplicity is obscured by their origin as worldsheet moduli space integrals (\ref{eq:definition Agn}).
This suggests the existence of a dual matrix model description.

\paragraph{Two-matrix model and topological recursion.} 
Double-scaled two-matrix integrals are familiar in the physics literature as the conjectured dual descriptions of the $(p,q)$ minimal string \cite{Kazakov:1986hu,Brezin:1989db,Douglas1991,Crnkovic:1989tn}. The dual of the $\mathbb{C}$LS is conceptually similar to this family of dualities but with important technical differences. The relevant class of two-matrix integrals is defined by integrating a pair of $N\times N$ Hermitian matrices weighted by suitable potentials together with a minimal coupling
\begin{equation}
    \langle \, \cdot \, \rangle = \!\int [\d M_1][\d M_2]\, (\, \cdot \,) \mathrm{e}^{-N\tr(V_1(M_1) + V_2(M_2) - M_1 M_2)} . \!
\end{equation}
The principal observables in such matrix models are resolvents, $R(x) \equiv \tr(x-M_1)^{-1}$, 
which exhibit a pole when $x$ coincides with an eigenvalue of $M_1$.
The relation between the matrix model and $\mathbb{C}$LS arises in a double-scaling limit, where we take $N\to \infty$ and zoom in on a particular region of the eigenvalue distribution of the matrices. Connected products of resolvents admit a topological expansion
\begin{equation}\label{eq:resolvent topological expansion}
    \Big\langle \prod_{j=1}^n R(x_j) \Big\rangle_{\text{c}} = \sum_{g=0}^\infty R_{g,n}(x_1,\ldots, x_n)\, \mathrm{e}^{S_0(2-2g-n)}\, ,
\end{equation}
where $\mathrm{e}^{S_0}$ is a proxy for $N$ held fixed in the double-scaling limit, corresponding to the inverse string coupling in \eqref{eq:sum over genera}.

The multi-resolvents $R_{g,n}(\boldsymbol{x})$ are multi-valued functions with branch cuts supported on the eigenvalue distribution of the matrix. They are thus naturally defined on a multi-sheeted cover of the complex plane with cuts on the real axis, which defines a Riemann surface known as the spectral curve. In the cases of interest, we can find a rational parameterization $(\xx(z),\yy(z)) \in \CC^2$ for $z \in \CC$. The multi-resolvents (\ref{eq:resolvent topological expansion}) are determined recursively in terms of the spectral curve via \emph{topological recursion} \cite{Eynard:2007kz,Eynard:2002kg,Chekhov:2006vd}, which efficiently solves the loop equations of the matrix model. To formulate topological recursion one views the resolvents as differentials on the spectral curve 
\be 
\omega_{g,n}(\boldsymbol{z})=R_{g,n}(\xx(\boldsymbol{z}))\,  \d \xx(z_1) \cdots \d \xx(z_n)~, \label{eq:resolvent differentials}
\ee
except for $(g,n)=(0,1)$, $(0,2)$ where the definition is slightly modified.

\paragraph{The spectral curve.}
Our central claim is that $\mathbb{C}$LS is dual to a double-scaled two-matrix integral characterized by the following spectral curve
\begin{equation}\label{eq:CLS spectral curve}
    \mathsf{x}(z) = -2\cos(\pi b^{-1} \sqrt{z}), \quad \mathsf{y}(z) = 2\cos(\pi b \sqrt{z})\, ,
\end{equation}
where $z$ is a uniformizing coordinate. 
This spectral curve leads to richer resolvents than that of the $(p,q)$ minimal string, since it (i) is not algebraic, and (ii) exhibits infinitely many branch points ($z_m^*$ such that $\d \mathsf{x}(z_m^*) = 0$) together with (iii) infinitely many nodal singularities (points $z^\pm$ that are mapped to the same point on the spectral curve). 

From the spectral curve (\ref{eq:CLS spectral curve}) one may extract the leading density of eigenvalues of the first matrix
by computing the discontinuity of $R_{0,1}(x)$:
\begin{equation}
    \rho_0(x) = \frac{2}{\pi}\sinh(-\pi i b^2)\sin\left(-ib^2 \cosh^{-1}(\tfrac{x}{2})\right)\, . \label{eq:density of states}
\end{equation}
This eigenvalue density is manifestly positive in a neighborhood of the edge $x = 2$, however it oscillates on sufficiently large scales, rendering the non-perturbative definition subtle as discussed below.

\paragraph{General string amplitudes.} The dictionary between the differentials $\omega_{g,n}$ and the string amplitudes $\mathsf{A}_{g,n}^{(b)}$ takes the form
\be 
\omega_{g,n}(\boldsymbol{z})= \int_0^\infty \prod_{j=1}^n \frac{-2\pi i p_j \d p_j \exp(2\pi i p_j \sqrt{z_j}) \d z_j}{\sqrt{z_j}}\, \mathsf{A}_{g,n}^{(b)}(\boldsymbol{p})~, \label{eq:Agn omegagn relation}
\ee
which can be inverted via an inverse Laplace transform.

One can use the relation between the quantities $\omega_{g,n}$ and intersection numbers on compactified moduli space $\bM_{g,n}$ to derive a general expression for $\mathsf{A}_{g,n}^{(b)}$,
\begin{align}
    \mathsf{A}_{g,n}^{(b)}(\boldsymbol{p}) &=\sum_{\Gamma \in \mathcal{G}_{g,n}^\infty}\frac{1}{|\text{Aut}(\Gamma)|}\int'\!\! \prod_{e \in \mathcal{E}_\Gamma} (-2 p_e \, \d p_e)\nonumber\\
    &\quad\times\prod_{v \in \mathcal{V}_\Gamma} \left(\frac{b(-1)^{m_v}}{\sqrt{2}\sin(\pi m_v b^2)}\right)^{2g_v-2+n_v}\nonumber\\
 &\quad\times  \prod_{j \in I_v} \sqrt{2} \sin(2\pi m_v b p_j) \mathsf{V}^{(b)}_{g_v,n_v}(i\boldsymbol{p}_v)~. \label{eq:Agn stable graphs}
\end{align}
The sum runs over all possible degenerations of the worldsheet as nodal surfaces. They are combinatorially labelled by stable graphs whose vertices $v \in \mathcal{V}$ indicate the different stable components of the surface of type $(g_v,n_v)$ while $e \in \mathcal{E}$ runs over the internal edges to which we assign a momentum $p_e$ that we integrate out. We also assign a color $m_v \in \mathbb{Z}_{\ge 1}$ to each vertex over whose choice we also sum. The quantity $\mathsf{V}_{g,n}^{(b)}(i\boldsymbol{p})$ is a polynomial of degree $3g-3+n$ in $p_j^2$ which remarkably is precisely the analytic continuation of the corresponding string amplitude of the Virasoro minimal string, which constitutes a quantum deformation of the Weil-Petersson volume \cite{Collier:2023cyw}. 
Finally, the prime on the integral denotes integration from $0$ to $\infty$ in the complex plane, computed by decomposing the trigonometric functions into exponentials and discarding divergent contributions without exponentials. 
We interpret the stable graphs, corresponding to degenerations of the worldsheet, as Feynman diagrams of the closed string field theory of $\mathbb{C}$LS in a suitable gauge, with the quantum volumes $\mathsf{V}_{g,n}^{(b)}$ as the string vertices.

\paragraph{Checks.} The expression \eqref{eq:Agn stable graphs} can be explicitly written out for low $(g,n)$, but the number of stable graphs grows rapidly for large $(g,n)$. For $(g,n)=(0,3)$, one can directly verify agreement with the worldsheet definition \eqref{eq:definition Agn} which does not involve any moduli integration. For $(g,n)=(0,4)$ and $(1,1)$, the expression \eqref{eq:Agn stable graphs} agrees with the result from the analytic bootstrap. For higher $(g,n)$, we proved that \eqref{eq:Agn stable graphs} has the correct poles \eqref{eq:general pole locations} and discontinuities as computed from the worldsheet, obeys the symmetries \eqref{eq:Agn symmetries} and satisfies the dilaton equation \eqref{eq:dilaton equation}. The duality symmetry \eqref{eq:Agn duality symmetry} is very non-trivial and is related to the $x-y$ symmetry of topological recursion \cite{Eynard:2007nq}.

\paragraph{Recursion relation.} The topological recursion of $\omega_{g,n}$ can be translated to the string amplitudes $\mathsf{A}_{g,n}^{(b)}$ via the relation \eqref{eq:Agn omegagn relation}. This leads to a recursion relation similar to Mirzakhani's recursion relation for the Weil-Petersson volumes \cite{Mirzakhani:2006fta}, except for a non-trivial sphere three-point amplitude $\mathsf{A}_{0,3}^{(b)}$ and a deformed integral kernel $\mathsf{H}_b$, which also appeared previously in \cite{Collier:2023cyw}.

\section{Worldsheet boundaries and non-perturbative effects}\label{sec:boundaries and non-perturbative effects} 
The closed string amplitudes (\ref{eq:definition Agn}) are contrived observables from the perspective of the dual matrix model. More natural observables are thermal partition functions (which admit a holographic interpretation in terms of a 2d gravity path integral on surfaces with asymptotic boundaries) or resolvent differentials (\ref{eq:resolvent differentials}) associated with one of the matrices. We denote such more general observables by $\mathsf{Z}^{(b)}_{g,n}(\boldsymbol{\Psi})$, where $\boldsymbol{\Psi}=(\Psi_1,\ldots,\Psi_n)$ is a set of boundary conditions tied to the specific observable.

\paragraph{Gluing conformal boundaries.}
The computation of $\mathsf{Z}^{(b)}_{g,n}(\boldsymbol{\Psi})$ can be encapsulated within a broader framework involving conformal boundary conditions for the string worldsheet CFT, labelled by a wavefunction $\Psi(p^+,p^-)$ in the product of the two Liouville theories. 
The computation of $\mathsf{Z}_{g,n}^{(b)}(\boldsymbol{\Psi})$ can then be reduced to the computation of $\mathsf{A}_{g,n}^{(b)}$.
This is in principle true for any critical worldsheet string theory but takes a remarkably simple form in the context of 2d or minimal string theories, where the space of physical states is particularly simple. 
In the case of $\CC$LS, the relation reads
\begin{equation}\label{eq:general partition function}
\mathsf{Z}^{(b)}_{g,n}(\boldsymbol{\Psi}) = 
\!\int\!\prod_{j=1}^n (-2 p_j \, \d p_j \, \mathsf{Z}^{(b)}_{\text{tr}}(\Psi_j,p_j) ) \, \mathsf{A}_{g,n}^{(b)}(\boldsymbol{p}) ~. 
\end{equation} 
Thus, all information in this broader class of observables $\mathsf{Z}^{(b)}_{g,n}(\boldsymbol{\Psi})$ is entirely captured by the elementary observable $\mathsf{Z}^{(b)}_{\text{tr}}(\Psi, p)\equiv\mathsf{Z}^{(b)}_{0,2}(\Psi, p)$.
This corresponds to the once-punctured disk diagram with $\Psi$ boundary condition, which is simply proportional to the on-shell wavefunction $\Psi(p^+=p,p^-=ip)$. In the context of 2d quantum gravity, this is commonly referred to as the ‘trumpet’ partition function \cite{Saad:2019lba}.
\eqref{eq:general partition function} means that a conformal boundary can be interpreted as a particular smearing of bulk vertex operators of the string worldsheet theory. 
Different worldsheet boundary conditions which lead to the same trumpet partition function are `BRST-equivalent' \cite{Seiberg:2003nm}.

Examples of trumpet partition functions of particular physical significance (a) reduce $\mathsf{Z}^{(b)}_{g,n}(\boldsymbol{\Psi})$ to the string amplitudes \eqref{eq:definition Agn}, (b) relate string amplitudes to matrix resolvent differentials \eqref{eq:resolvent differentials}, (c) compute thermal partition functions of the matrix model, and (d) contribute to the non-perturbative effects mediated by ZZ-instantons:
\begin{subequations}
\begin{align}
    \mathsf{Z}^{(b)}_{\text{tr}}(q,p)&= \mathsf{A}_{0,2}^{(b)}(q,p) = -\frac{\delta(p-q)}{2p} ~~ \text{(trivial)} \, , \\
    \mathsf{Z}^{(b)}_{\text{tr}}(z,p)&= -\pi \sin(2\pi \sqrt{z}p) \, \frac{\d z}{\sqrt{z}} ~~~ \text{(resolvent)} \, , \\
    \mathsf{Z}^{(b)}_{\text{tr}}(\beta,p)&= \frac{2b}{\pi} \sin(2\pi bp) K_{2bp}(2\beta) ~ \text{(thermal)} \, , \\
    \mathsf{Z}^{(b)}_{\text{tr}}((r,s),p)&= \frac{2\sin(2\pi rbp)\sin(2\pi sb^{-1}p)}{p} ~~ \text{(ZZ)} \, . \label{eq:ZZ trumpet}%
\end{align}
\end{subequations}

\paragraph{Non-perturbative effects.} The genus expansion \eqref{eq:sum over genera} is an asymptotic series that admits non-perturbative corrections of order $\exp(\# g_\text{s}^{-1})=\exp(\# \exp S_0)$.
These corrections can be systematically computed from open worldsheets with ZZ-boundary conditions \cite{Polchinski:1994fq,Balthazar:2019rnh,Balthazar:2019ypi}. 
A complete basis of ZZ-instanton boundaries is spanned by the $\text{ZZ}_{(r,s)} \times \text{ZZ}_{(1,1)}$ boundary condition \cite{Zamolodchikov:2001ah} in the two Liouville factors with $r,s \in \ZZ_{\ge 1}$. 
In particular, the tension of the ZZ-instanton is computed from the empty disk diagram, which can be extracted from the dilaton equation \eqref{eq:dilaton equation} and the ZZ-trumpet \eqref{eq:ZZ trumpet},
\be 
T_{r,s}^{(b)}=8\mathrm{e}^{S_0} (-1)^{r+s} \, \frac{\sin( \pi r b^2 )\sin( \pi s b^{-2} )}{b^{-2}-b^2} ~.
\ee 
Notably, $T_{r,s}^{(b)}$ is purely imaginary. 
Therefore, these non-perturbative effects of order $\exp(-T_{r,s}^{(b)})$ are highly oscillatory, instead of exponentially suppressed. This also implies via resurgence that the large order behaviour of $\mathsf{A}_{g,n}^{(b)}$ as $g \to \infty$ is alternating and the effective string coupling is imaginary, see also \cite{Cotler:2024xzz}.

The technology of gluing trumpets can be applied to ZZ-instantons. 
An exception is the cylinder with identical instanton boundary conditions, where the integral in \eqref{eq:general partition function} diverges due to the breakdown of Siegel gauge in the open string zero-mode sector. 
This divergence can be resolved using string field theory \cite{Sen:2020cef,Sen:2021qdk,Eniceicu:2022nay}. 

On the two-matrix model side, non-perturbative corrections are computed by integrating a single eigenvalue pair over a steepest descent contour of the effective potential, with saddle points corresponding to the worldsheet ZZ-instantons. 
The technology for saddle point evaluation in matrix models is well-established \cite{Daul:1993bg, Eniceicu:2022dru} and reproduces the leading non-perturbative corrections to $\mathsf{A}_{n}^{(b)}(S_0;\boldsymbol{p})$ (\ref{eq:sum over genera})
as computed diagrammatically from the string worldsheet in arbitrary multi-instanton sectors. 
This analysis probes non-perturbative aspects of the duality.

\paragraph{Non-perturbative completion.} On the matrix model side, the locations of the instanton saddles coincide with the zeros of the density of states \eqref{eq:density of states}. This indicates that \eqref{eq:density of states} can only be trusted up to its first zero at $E_0=2 \cos(\pi b^{-2})$, after which the integration contour over the eigenvalue pairs is deformed into the complex plane. Different choices for this contour result in inequivalent non-perturbative completions. 
This effectively yields a model whose density of states is supported on a finite interval, reminiscent of \cite{Berkooz:2018jqr,Narovlansky:2023lfz,Verlinde:2024zrh}.

\section{Gravitational path integral} \label{sec:quantum gravity}
\paragraph{Sine dilaton gravity.} A field redefinition recasts the worldsheet action of $\CC$LS (\ref{eq:Liouville squared worldsheet definition}) as the following dilaton gravity action \cite{Mertens:2020hbs,Collier:2023cyw,Goel:2022pcu,Blommaert:2023wad,Blommaert:2024ydx} 
\begin{equation}\label{eq:S+Ss} 
S_{\mathrm{dil}}[\Phi,g] = \frac{i}{2b^2}\int_{\Sigma_{g,n}} \!\!\!\d^2 x\sqrt{g}\, \big(\Phi \mathcal{R} + \frac{1}{\pi}\sin(2\pi \Phi)\big)~,
\end{equation}
where $\Phi$ is the dilaton and we omit boundary contributions. 
The reciprocal of the prefactor can be interpreted as the gravitational coupling constant, with a semiclassical limit as $-ib^2\rightarrow 0$. 
The $\mathbb{C}$LS string amplitudes are reinterpreted as a gravitational path integral 
\begin{equation} \label{eq:Agn PI} 
\mathsf{A}^{(b)}_{g,n}(\mathbf{p})\equiv\int \frac{[\mathcal{D}g][\mathcal{D}_g \Phi]}{\rm vol_{\rm diff}}\, \mathrm{e}^{-S_{\mathrm{dil}}[\Phi,g]}\mathcal{V}_{p_1}\cdots \mathcal{V}_{p_n}~,
\end{equation}
with the vertex operators defined previously.
Solutions to the equations of motion of (\ref{eq:S+Ss}) come in a family of alternating dS and AdS vacua,
\begin{subequations}
\begin{align} 
   \mathcal{R}_* &= 2(-1)^{m+1} +4\pi \sum_{j=1}^n(1-2b p_j)\delta^2(\xi- \xi_j) \\
 \Phi_* &= \frac{m}{2}~,\quad m\in \mathbb{Z}~,
    \end{align}\label{eq: eom saddles}%
\end{subequations}
somewhat reminiscent of \cite{Anninos:2017hhn, Anninos:2022hqo}. Here $\xi_j$ are the locations of the vertex operator insertions. We have thus found an infinite number of classical saddles with constant dilaton labelled by an integer $m\in \mathbb{Z}$. For even $m$, the Ricci scalar is negative, whereas for odd $m$ we obtain a positive scalar curvature. The sine dilaton gravity theory (\ref{eq:S+Ss}) thus admits a family of alternating anti-de Sitter and de Sitter vacua.
The dS solutions necessitate a $(-,-)$ signature of the metric, similar to the dS saddles of \cite{Cotler:2024xzz}. $m$ can be identified with the color $m_v$ appearing in the exact answer \eqref{eq:Agn stable graphs}, which suggests that only the saddles $m\geq 1$ contribute to the path integral \eqref{eq:Agn PI}. 
\eqref{eq:Agn stable graphs} also suggests that we should consider more general saddles consisting of degenerated Riemann surfaces connected at nodes whose different components $v \in \mathcal{V}$ can be in different vacua $m_v \in \ZZ_{>0}$. These represent transition amplitudes between universes with different cosmological constants. 
Evaluating the path integral \eqref{eq:Agn PI} beyond the leading saddle we uncover the same fluctuation theory as in the Virasoro minimal string \cite{Collier:2023cyw}.
Together this provides a path integral interpretation of the sum over stable graphs (\ref{eq:Agn stable graphs}).

\section{Outlook}
We conclude with an outlook on the relation of the complex Liouville string and 3d de Sitter quantum gravity. 

In \cite{paper5} we carefully quantize the gravitational phase space of dS$_3$ quantum gravity on a compact, hyperbolic Cauchy slice $\Sigma_{g,n}$. We motivate that the Liouville correlator $\Psi^{(b)}_{g,n}=\langle V_{p_1} \cdots V_{p_n} \rangle_{\Sigma_g}$ with central charge $c\in 13+i\mathbb{R}$ is a natural wavefunction of the universe at $\mathcal{I}^+$. It corresponds to an expanding spacetime with a big bang singularity in the past, and its norm is precisely computed by the string amplitude $\mathsf{A}^{(b)}_{g,n}$ of the complex Liouville string. We interpret the norm of the wavefunction as defining cosmological correlators of massive particles in dS$_3$, integrated over the metric at $\mathcal{I}^+$. This constitutes a first piece of evidence toward a microscopic realization of dS$_3$ in terms of the matrix integral dual of the complex Liouville string. The second piece of evidence ties to the Gibbons-Hawking de Sitter entropy conjecture \cite{Gibbons:1976ue,Gibbons:1977mu}. According to this, the cosmological event horizon surrounding an observer in dS captures an entropy, which is computed by the logarithm of the Euclidean gravity path integral on the sphere (the Euclidean continuation of de Sitter). We observe in \cite{paper5} that truncating the eigenvalue density (\ref{eq:density of states}) at its first zero, the Gibbons-Hawking entropy can be reproduced microscopically by counting the number of eigenvalues in the dual matrix model.

\section{Acknowledgements}
We would like to thank Dionysios Anninos, Aleksandr Artemev, Mattia Biancotto, Matthias R.\ Gaberdiel, Davide Gaiotto, Alessandro Giacchetto, Victor Godet, Victor Gorbenko, Clifford V. Johnson, Shota Komatsu, Adam Levine, Raghu Mahajan, Juan Maldacena, Alex Maloney, Marcos Mari\~no, Nico Valdes-Meller, Mehrdad Mirbabayi, Miguel Montero, Nikita Nekrasov, Hirosi Ooguri, Boris Post, Maximilian Schwick, Ashoke Sen, Douglas Stanford, J\"org Teschner, Cumrun Vafa, Herman Verlinde, Edward Witten and Xi Yin for discussions. 
SC, LE and BM thank l’Institut Pascal
at Universit\'e Paris-Saclay, with the
support of the program ``Investissements d’avenir'' ANR-11-IDEX-0003-01, 
and SC and VAR thank the Kavli Institute for Theoretical Physics (KITP), which is supported in part by grant NSF PHY-2309135, for hospitality during the course of this work. 
VAR is supported in part by the Simons Foundation Grant No. 488653, by the Future Faculty in the Physical Sciences Fellowship at Princeton University, and a DeBenedictis Postdoctoral Fellowship and funds from UCSB. 
SC is supported by the U.S. Department of Energy, Office of Science, Office of High Energy Physics of U.S. Department of Energy under grant Contract Number DE-SC0012567 (High Energy Theory research), DOE Early Career Award  DE-SC0021886 and the Packard Foundation Award in Quantum Black Holes and Quantum Computation. BM gratefully acknowledges funding provided by the Sivian Fund at the Institute for Advanced Study and the National Science Foundation with grant number PHY-2207584.

\bibliographystyle{apsrev4-1}
\bibliography{letterbib}
\end{document}